\documentclass[twocolumn]{aa}  

\newcommand{\htwo}{H$_2$}

\newcommand{\htwop}{H$_2^+$}
\newcommand{\hthreep}{H$_3^+$}
\newcommand{\hp}{H$^+$}
\newcommand{\hep}{He$^+$}
\newcommand{\hehp}{HeH$^+$}

\newcommand{\cotwo}{CO$_2$}

\newcommand{\hi} {H {\sc i}}

\newcommand{\hei} {He {\sc i}}
\newcommand{\heii}{He {\sc ii}}

\newcommand{\lalpha}{Lyman-$\alpha$}

\newcommand{\eminus} {e$^-$}

\usepackage{graphicx}

\usepackage{txfonts}
\usepackage{xcolor}
\usepackage{natbib}
\usepackage{ulem}
\usepackage{float}

\usepackage{hyperref}

\begin{document}

   \title{Modelling helium in exoplanet atmospheres. \\
   A revised network with   
   photoelectron-driven processes.}

   \author{A. Garc\'ia Mu\~noz
          \inst{1}
          }

   \institute{
   Universit\'e Paris-Saclay, Universit\'e Paris Cit\'e, CEA, CNRS, AIM, 91191, Gif-sur-Yvette, France
    \email{antonio.garciamunoz@cea.fr}
    }
 
  \abstract 
  {The {\hei} line at 1.08 $\mu$m is a valuable tracer of atmospheric escape in exoplanet atmospheres.
  }
  {We expand past networks used to predict the
  absorbing He(2$^3S$) by including, firstly, processes that involve {\htwo} and some molecular ions and, secondly, the interaction of photoelectrons with the atmosphere.
  }
  {We survey the literature 
  on the chemical-collisional-radiative processes that govern the production-loss of He(2$^3S$). 
We simulate the atmospheric outflow from the Neptune-sized GJ 436 b by coupling 
a hydrodynamic model that solves the bulk properties of the gas and a Monte Carlo model that tracks the energy degradation of the photoelectrons.}
  {We identify Penning ionization of H as a key He(2$^3S$) loss process at GJ 436 b and update its rate coefficient to a value consistent with the most recent available cross sections. The update affects notably the predicted strength of the {\hei} line. 
  For GJ 436 b, photoelectron-driven processes (mainly ionization and excitation) modify the He(2$^3S$) population in layers too deep to affect the in-transit spectrum. The situation might be different for other atmospheres though.
  The spectral energy distribution of GJ 436 has a strong effect on the predicted in-transit signal. 
  The published non-detections of the {\hei}  
  line for GJ 436 b are reasonably consistent with our model predictions for a solar-metallicity atmosphere  
  when the model adopts a recently proposed spectral energy distribution.  
  }
{The interpretation of the {\hei} line at 1.08 $\mu$m is model-dependent. Our revised network provides a general framework to extract more robust conclusions from measurements of this line, especially in atmospheres where {\htwo} remains abundant to high altitudes. We will explore additional, previously-ignored processes in future work.}


   \titlerunning{New aspects in the modelling of the {\hei} line at 1.08 $\mu$m}
   \maketitle

\section{Introduction}

Atmospheric escape plays a key role in the evolution of exoplanets \citep{owen2019,beanetal2021}. 
Diagnostic lines that trace the escaping gas and can be probed with current technology, such as those in the {\hi} Balmer series 
\citep{yanhenning2018}
or the {\hei} line at 1.08 $\mu$m, are advancing enormously our understanding of atmospheric escape. The history of the latter is particularly interesting. 
Modelling by \citet{seagersasselov2000} and \citet{turneretal2016} suggested that it was detectable at some close-in exoplanets. It took however almost two decades for the first detection in the in-transit spectrum of an exoplanet to be reported \citep{spakeetal2018}. 
Since then, the {\hei} line at 1.08 $\mu$m has been identified at $\sim$20 exoplanets and remains a priority in spectroscopic surveys \citep{vissapragadaetal2022,fossatietal2023,guilluyetal2024,massonetal2024,sanz-forcadaetal2025}.
As observations of this line continue, it becomes more important to assess its diagnostic value and how its strength is affected by the planet-star properties.
\\

\citet{oklopcichirata2018} laid out a network of processes that populate the  metastable state absorbing in the He(2$^3S$){+}{$h\nu$ (1.08 $\mu$m)}$\rightarrow$He(2$^3P$) line. 
In short, photoionization by energetic stellar photons creates {\hep} ions that recombine with the thermal electrons.
A fraction of the recombinations end up as He(2$^3S$) which, not being connected with the ground state through dipole emission, builds up to potentially detectable amounts. 
Subsequent models adopted this network with little variations \citep[e.g.][]{oklopcic2019,czeslaetal2022,dossantosetal2022,
yanetal2022,lamponetal2023,
rumenskikhetal2023,biassonietal2024,
allanetal2024,schreyeretal2024,
ballabioowen2025}. 
Taking a step back, we
aim to assess the completeness of the network. 
In particular, in this first work, 
we expand the network to accommodate interactions with {\htwo} and some molecular ions, and
explore to what extent photoelectron-driven processes affect directly or indirectly the He(2$^3S$) population. In later work, we will explore other aspects of the modelling such as the importance of in-detail radiative transfer in the lines and the continuum, or the role of chemical reactions with a variety of elements.
\\

We define by photoelectrons or primaries the electrons 
formed from photoionization, and by secondaries the electrons formed as the primaries slow down in ionizing collisions 
with atoms and molecules. 
A few works have explored the role of photoelectrons in the energy budget and chemistry of exoplanets \citep{cecchi-pestellinietal2009,shematovichetal2014,
garciamunoz2023_aa,
gilletetal2023,loccietal2024}. 
Their effect on diagnostic lines remains however largely unexplored \citep{garciamunoz2023_icarus,gilletetal2025}. 
To our knowledge, and save for some qualitative statements in \citet{garciamunoz2023_icarus}, 
there has been no quantitative study of the effect of photoelectrons on the {\hei} line at 1.08 $\mu$m.
\\

We focus on GJ 436 b, a 
Neptune-sized exoplanet orbiting an M2.5V-type star.
The planet has been a recurrent target of 
observations 
to characterize its atmosphere and of modelling for their interpretation 
\citep[e.g.][]{madhusudhanseager2011,mosesetal2013,knutsonetal2014,
morleyetal2017,grasseretal2024,mukherjeeetal2025}. 
Given its low density and 
equilibrium temperature $T_{\rm{eq}}${$\sim$}
700 K, the atmospheric layers probed 
by visible-infrared radiation are likely dominated by {\htwo}. 
Photodissociaton will lead into an H-atom-dominated gas at some  high altitude. 
It is unclear whether the transition from {\htwo}- to H-dominated has implications on other aspects of the atmospheric chemistry, which remains a path to be explored.
The evidence at GJ 436 b for any molecules, e.g. H$_2$O, CO, {\cotwo}, CH$_4$ or {SO$_2$} that have been found at other warm exoplanets, remains weak
\citep{grasseretal2024,mukherjeeetal2025}. The non-detection of molecular features is attributed to the atmosphere having a high metallicity and therefore a small scale height, to the occurrence of high-altitude clouds or to a combination of both conditions. 
Identifying the main reason remains an open problem that, once solved, should help better understand the generality of exoplanet atmospheres. 
As other Neptune-sized planets such as HAT-P-11 b and GJ 3470 b, GJ 436 b is 
enshrouded by a cloud of H atoms escaping the planet and that can be probed by {\hi} {\lalpha} absorption spectroscopy \citep{kulowetal2014,ehrenreichetal2015,bourrieretal2018,
ben-jaffeletal2022}. Unlike them, 
the attempts to detect the {\hei} line at 1.08 $\mu$m at GJ 436 b
have so far failed \citep{nortmannetal2018,guilluyetal2024,massonetal2024}. The non-detection of the {\hei} line
is intriguing because the available models predict a strong absorption signal 
\citep{oklopcichirata2018,dossantosetal2022,rumenskikhetal2023}. 

\section{General formulation}

To simulate the escaping atmosphere, we solve the equations of mass, momentum and energy conservation for the multi-species gas in a spherical-shell geometry.
The helium signal peaks relatively close to the planet, and therefore this simplified geometry should be adequate to model the absorbing region.
The numerical model  
builds upon methods presented previously \citep{garciamunoz2007b,garciamunozetal2021}, and currently under re-development to incorporate additional disequilibrium processes (Garc\'ia Mu\~noz, \textit{in prep.}). In what follows, we give a brief account of the helium modelling.
\\

The model solves the mass conservation equation:
\begin{equation}
\frac{\partial n_s}{\partial t} + \nabla \cdot \Phi_s
 =  \frac{\delta n_s}{\delta t},
\label{mass_conservation_eq}
\end{equation}
where $n_s$ [cm$^{-3}$] is the number density of the species (or particle type) specified by $_s$, $\Phi_s$ [cm$^{-2}$s$^{-1}$] is the flux, and $\delta n_s/\delta t$ [cm$^{-3}$s$^{-1}$] 
is the production or loss rate (if $>$0 or $<$0, respectively) due to chemical, collisional and radiative (Ch-C-R, hereafter) processes.
We omit for now  
photoexcitation between bound states driven by absorption of radiation. The 
implications of this simplification
will be investigated in future work on the detailed treatment of radiative transfer.
In its current implementation, the network includes about 200 Ch-C-R processes.
The momentum conservation equation is treated in a standard way. 
Energy conservation is formulated by consistently tracking the radiative exchanges (for bound-bound, bound-free/free-bound and free-free interactions) of the gas with its surroundings \citep{garciamunozschneider2019}.  \\

We focus on purely hydrogen-helium 
atmospheres made of {\eminus} (thermal electrons), H($i$), {\htwo}, 
{\hp}, {\htwop}, {\hthreep}, He($i$),  {\hep} and {\hehp}. Index
$i$ specifies the electronic excitation state. 
The hydrogen chemistry implemented in the model is relatively standard \citep{garciamunozschneider2019,garciamunozetal2021}, and will not be discussed further. 
The He($i$) states are listed 
in Table \ref{states_table}. 
The  He(2$^3S$) and He(2$^1S$) states are metastable. 
 Truncating the He($i$) atom model at $i${$\le$}5 is a reasonable trade-off between expediency and an accurate representation of the cascade that forms upon radiative recombination. 
We omit the doubly-charged ion He$^{2+}$ because its density is small under most conditions. A separate form of Eq. \ref{mass_conservation_eq} is solved for each species.\\

We made a significant effort to revise the Ch-C-R processes relevant to the species listed above, in particular those that contain helium. 
For example, our network incorporates a variety of processes in which they interact with {\htwo} and 
the {\htwop}, {\hthreep} and {\hehp} molecular ions that appear as by-products of the {\htwo} chemistry. This is essential for simulating the warm atmospheres of Neptune- and sub-Neptune-sized planets 
that are likely to retain {\htwo} up to high altitudes. 
Next, we elaborate on our choices for the Ch-C-R processes participated by helium-bearing species.
\\

Photoionization is considered for the He ground and metastable states, with cross sections from the NORAD database \citep{nahar2010,nahar2020}. They are presented in Fig. \ref{xs_he_fig}. The structure at 100-400 {\AA} is real, and arises from resonances. For the inverse process of 
radiative recombination, {\hep}+{\eminus}{$\rightarrow$}He($i$)+{$h \nu$}, 
we calculated ourselves effective rate coefficients 
$\alpha_{i,\rm{eff}}$ [cm$^3$s$^{-1}$]. Each
$\alpha_{i,\rm{eff}}$ includes the contribution $\alpha_{i}$ from direct radiative recombination into state $i$
 plus the contribution from the cascade that forms from the higher-energy states not explicitly resolved in our atom model and that eventually radiate into state $i$. 
Formally, 
$\alpha_{i,\rm{eff}}$=$\alpha_{i}$+$\sum_{j>5}${$\alpha_{j}$}{$p_{ji}$}, where {$p_{ji}$}=$A_{ji}$/$\sum_{i,i\ne 1} A_{ji}$ is the probability that emission from state $j$ leads into state $i$ under the assumption that line opacity prevents the radiative decay into the ground state. 
For the cascade, 
we borrowed $\alpha_i$ and $A_{ji}$ from NORAD. 
For states with principal quantum number {$>$}10, NORAD reports the joint contribution of recombination into singlets and triplets. We assumed a 1:3 partitioning between them, and assigned their rates into the states with principal quantum number =10.
Table \ref{radrecom_table} shows the $\alpha_{i,\rm{eff}}$ at a few temperatures. 
Recombination into the higher-energy (lower-energy) states becomes prevalent at low (high) temperatures. 
For implementation, we fitted 
$\alpha_{i,\rm{eff}}$ to 
analytical expressions between 100 and 30,000 K.
These fits, and others described later, are made available in a SI file. 
\\

We took the transition probabilities $A_{ji}$ 
for bound-bound transitions from \citet{wiesefuhr2009}. 
Exceptionally, $A_{ji}$ for 
two-photon He(2$^1S$){$\rightarrow$}He(1$^1S)$+$h\nu_1$+$h\nu_2$ is   
from \citet{drake1986}. See Table \ref{Aji_table} for details. 
\\

We formed the rate coefficients for (de)excitation 
of the He($i$) states 
in collisions with thermal electrons from the effective collision strengths $\Upsilon_{ji}$($T$) 
of \citet{berringtonkingston1987} at temperatures 
$T$ between 1,000 and 5,000 K and \citet{brayetal2000} between 5,600 and 20,000 K. 
The latter calculations are expected to be more accurate. 
An accurate description of the low temperatures is 
critical though because the He($2^3S$) absorption often occurs in relatively cool layers of the atmosphere. 
Coincidentally or not, the $\Upsilon_{ji}$($T$) from both works 
for deexcitation of the He metastables
are consistent. Confirmation of the 
low-temperature $\Upsilon_{ji}$($T$) by new calculations is welcome to minimize the model uncertainties. Table \ref{em-collexc_table} shows the rate coefficients at a few temperatures. 
The collisions of charged particles, e.g. {\hp} or {\hep}, with the 
He($i$) states might conceivably contribute to their (de)excitation. Unfortunately, there seems to be no quantitative information on these collisions 
 at the relevant energies/temperatures. 
Calculations of this type are welcome.
\\

The loss of the He metastables in collisions with neutrals proceeds typically via ionization. Table \ref{penning_table} lists the dominant processes \citep{milleretal1972,westetal1975,prestoncohen1976}, 
as implemented here. 
For collisions with H and {\htwo}, we calculated ourselves the rate coefficients from the total ionization cross sections reported in the quoted references to ensure they are reliable from 200 to 10,000 K. For the partitioning of Penning:associative ionization (the latter leads to {\hehp}), we used an average 0.9:0.1.
We note that the literature contains different prescriptions for Penning ionization. 
For example, 
\citet{robergedalgarno1982} quote a
$T$-independent rate coefficient of 5$\times$10$^{-10}$ cm$^3$s$^{-1}$ 
for Penning ionization of H atoms,
whereas \citet{fassiaetal1998} propose 7.5$\times$10$^{-10}$($T$/300)$^{1/2}$ 
cm$^3$s$^{-1}$ for the same process. 
For a temperature $T$=5,000 K representative of exoplanet conditions, the two prescriptions differ by a factor of 6. 
Table \ref{hehp_table} lists additional processes that control the {\hehp} population.
\\

We consider the charge-exchange processes in Table \ref{chargeexchange_table}. 
They include the loss of He metastables in collisions with {\hp}, with rate coefficients from \citet{loreauetal2018} extrapolated to $T${$<$}3,000 K (J\'er\^ome Loreau, \textit{priv. comm.}).
For {\hep}+{\htwo}, we adopted a rate coefficient based on the \citet{schaueretal1989} experiment and assumed that the process leads to 
{\htwo} dissociation.
\\

\begin{figure}
   \centering
   \includegraphics[width=9.2cm]{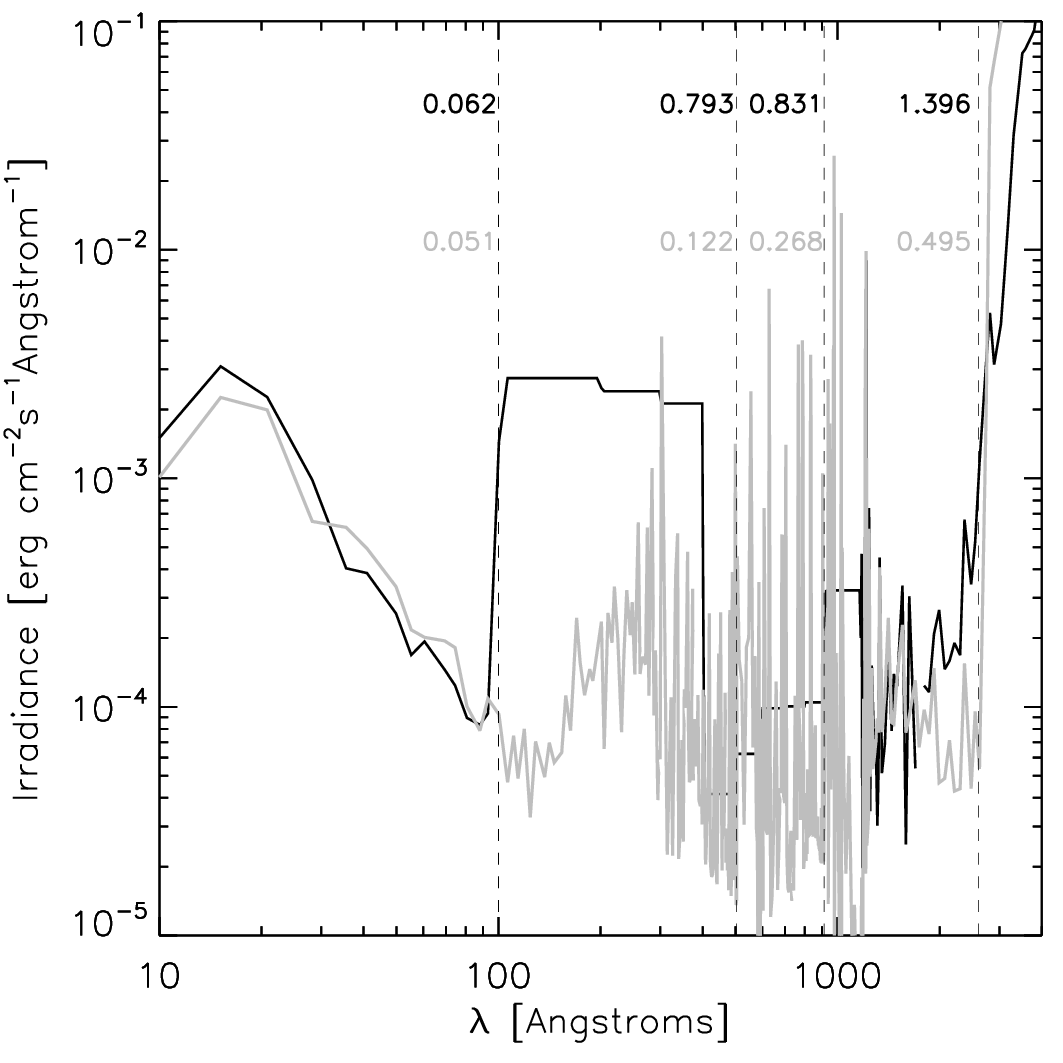}
      \caption{\label{SED_fig} 
      MUSCLES SED (black) and X-Exoplanets SED (grey) at 1 AU and the spectral resolution adopted here. The quoted numbers refer, from left to right, to the integration over the intervals  0{$\rightarrow$}100 {\AA}; 0{$\rightarrow$}503 {\AA}; 
      0{$\rightarrow$}911 {\AA};
      0{$\rightarrow$}2,600 {\AA}, in units of erg cm$^{-2}$s$^{-1}$.
      }
\end{figure}

The spectral energy distribution (SED) of the host star partly dictates the rate 
at which photoelectrons are released. 
Prescriptions of GJ 436's SED have been presented via the HAZMAT \citep{peacockeal2019}, 
MUSCLES \citep{franceetal2016} and 
X-Exoplanets \citep{sanz-forcadaetal2025}
programs. 
There are notable differences amongst them, and we refer to those references to understand their origin.
The HAZMAT SED includes no X-ray information and thus we omit it from our discussion. We repeated all our calculations with both the MUSCLES and the X-Exoplanets SED. 
We extended the X-Exoplanets SED above its original long-wavelength limit of 2,800 {\AA} by assuming blackbody emission from the star. The extension introduces an unphysical discontinuity in the SED. As our calculations for GJ 436 b show that photoionization does not dominate the He(2$^1S$) loss (the metastable that photoionizes at these wavelengths), we have not attempted an alternative reconstruction.
Figure \ref{SED_fig} shows both the MUSCLES and X-Exoplanets SED at the spectral resolution at which our model calculations are performed.
\\

The boundary conditions in the hydrodynamic model
that solves the atmospheric outflow 
are standard \citep[][]{garciamunoz2007b}. 
The lower boundary is set at a radial distance $R_{\rm{p}}$ that we take as the planet's optical radius. We fix there the pressure to $p$=1 dyn cm$^{-2}$, the temperature to $T_{\rm{eq}}$ and the volume mixing ratios to
values appropriate to solar composition 
(0.163 for He, a somewhat arbitrary 10$^{-3}$ for H, and 0 for the other species except {\htwo}, which makes the bulk of the gas). 
We explored the impact of setting the lower boundary at pressures as high as 100 dyn cm$^{-2}$. It was found to be relatively minor because for a temperate planet such as GJ 436 b the geometric size of the region between 100 and 1 
dyn cm$^{-2}$ is notably less than that of the layer masked by the {\hei} line core. 
\\

The hydrodynamic model is coupled to a Monte Carlo model,
with which the photoelectron-driven rates for excitation, dissociation and ionization are calculated
\citep{garciamunoz2023_icarus,garciamunozbataille2024}. Both models communicate with each other until a steady state is attained. 
In our experience, only a few Monte Carlo calculations are needed to describe accurately the  photoelectron effects.

\section{Photoelectron-driven processes}

Based on energy considerations, it is apparent that 
photoelectrons can drive numerous processes  that are inaccessible to the thermal electrons. 
Exploring this idea, 
\citet{garciamunoz2023_icarus} studied some aspects of the photoelectron-driven
production and loss of the H(2$s$) and H(2$p$) states that are probed via the {\hi} Balmer lines in the in-transit spectra of ultra-hot Jupiters \citep{yanhenning2018,cauleyetal2021,
czeslaetal2022}. That work found that for the exoplanet HAT-P-32 b: 
\textit{i}) 
Photoelectron-driven excitation of
 H ground state does not dominate the 
H(2$s$)/H(2$p$) formation where the {\hi} Balmer lines form;  \textit{ii}) The collisions of photoelectrons with H(2$s$)/H(2$p$), whether leading to deexcitation or ionization, affect negligibly the lifetimes of these excited states; 
\textit{iii}) Secondary electrons potentially modify the ionization balance in the atmosphere at altitudes where the {\hi} Balmer lines and the {\hei} line at 1.08 $\mu$m are formed. 
\citet{garciamunoz2023_icarus} did not solve the simultaneous hydrodynamic and photoelectron energy-degradation problems, and therefore did not quantify the implications of point \textit{iii}).
\\

The present study revisits and expands on  the above ideas, focusing on the {\hei} line at 1.08 $\mu$m at GJ 436 b. 
In particular, we aim to quantify 
the contribution of photoelectrons to:
\begin{itemize}
    \item Q1.  The ionization balance of an {\htwo}/H/He atmosphere, and how this propagates into the  He(2$^3S$) population. 
    \item Q2. The production of He(2$^3S$), either directly from the  He ground state or indirectly through excitation of higher-energy states that cascade through bound-bound radiation.
    \item Q3. The He(2$^3S$) loss through ionization or deexcitation.    
\end{itemize}

To address these questions, we newly included in the Ch-C-R network the relevant processes 
for photoelectron-driven 
(de)excitation, dissociation and 
 ionization of {\htwo}, H and He. The treatment of H and {\htwo} was described before \citep{garciamunoz2023_icarus,garciamunozbataille2024}.
The photoelectron-driven processes for He are listed in Table \ref{nonthermalchannels_table}.
The cross sections for He excitation and ionization are 
from \citet{ralchenkoetal2008}, and we
inferred the deexcitation ones from detailed balancing. Figure \ref{electronimpactxs_fig} shows them. Recent experiments by 
\citet{genevriez2017} confirm that the adopted He(2$^3S$) ionization cross sections are reliable.
\\

\begin{figure*}
   \centering
   \includegraphics[width=14.cm]{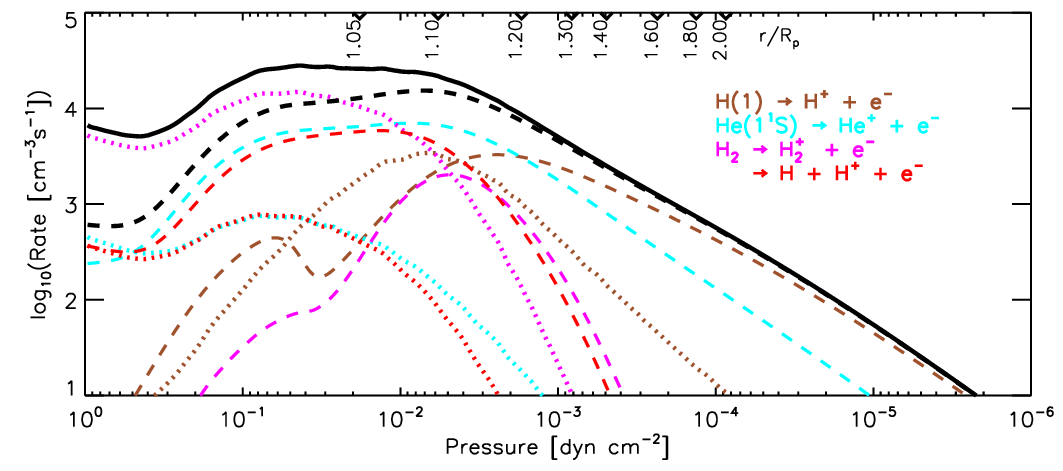}   
   \includegraphics[width=14.cm]{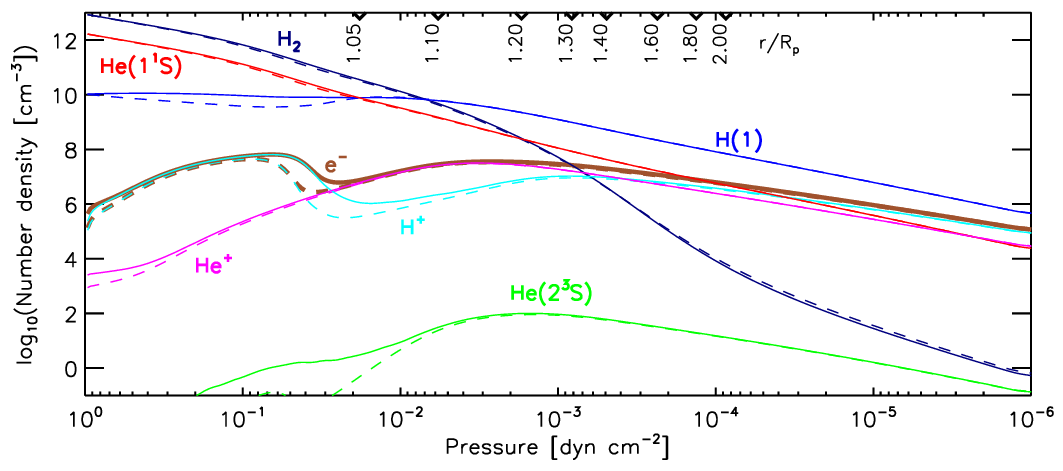}   
      \caption{\label{profiles_fig} 
      Top. 
      From the MUS-a and -b models, production
      rates of primary (dashed lines) and secondary (dotted lines)
      electrons. Colors separate the contributions by donor. The dashed black line refers to the primaries summed over all donors. 
      The solid black line refers to the total of primaries and secondaries summed over all donors. 
      Bottom. Number densities. Dashed and solid lines refer to MUS-a and -b, respectively.
      Each panel provides, based on the MUS-b model, the conversion between pressure and radial distance normalized to the planet's optical radius ($r$/$R_{\rm{p}}$; =1 at the model's lower boundary). 
}
\end{figure*}

\begin{figure}
   \centering
   \includegraphics[width=9.cm]{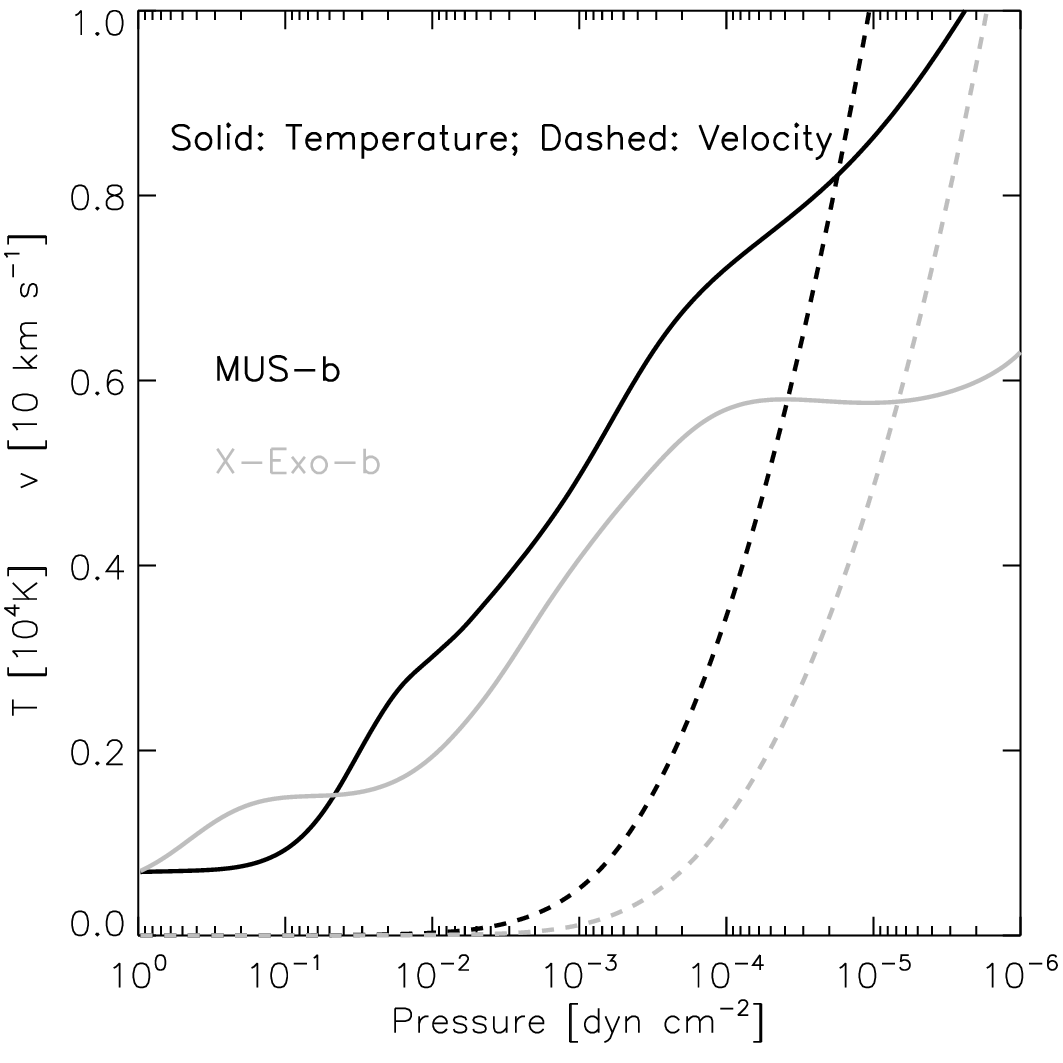}   
      \caption{\label{velotemp_fig} From the MUS-b and X-Exo-b models, temperature and velocity profiles.      
}
\end{figure}

\begin{figure}
   \centering
   \includegraphics[width=9.cm]{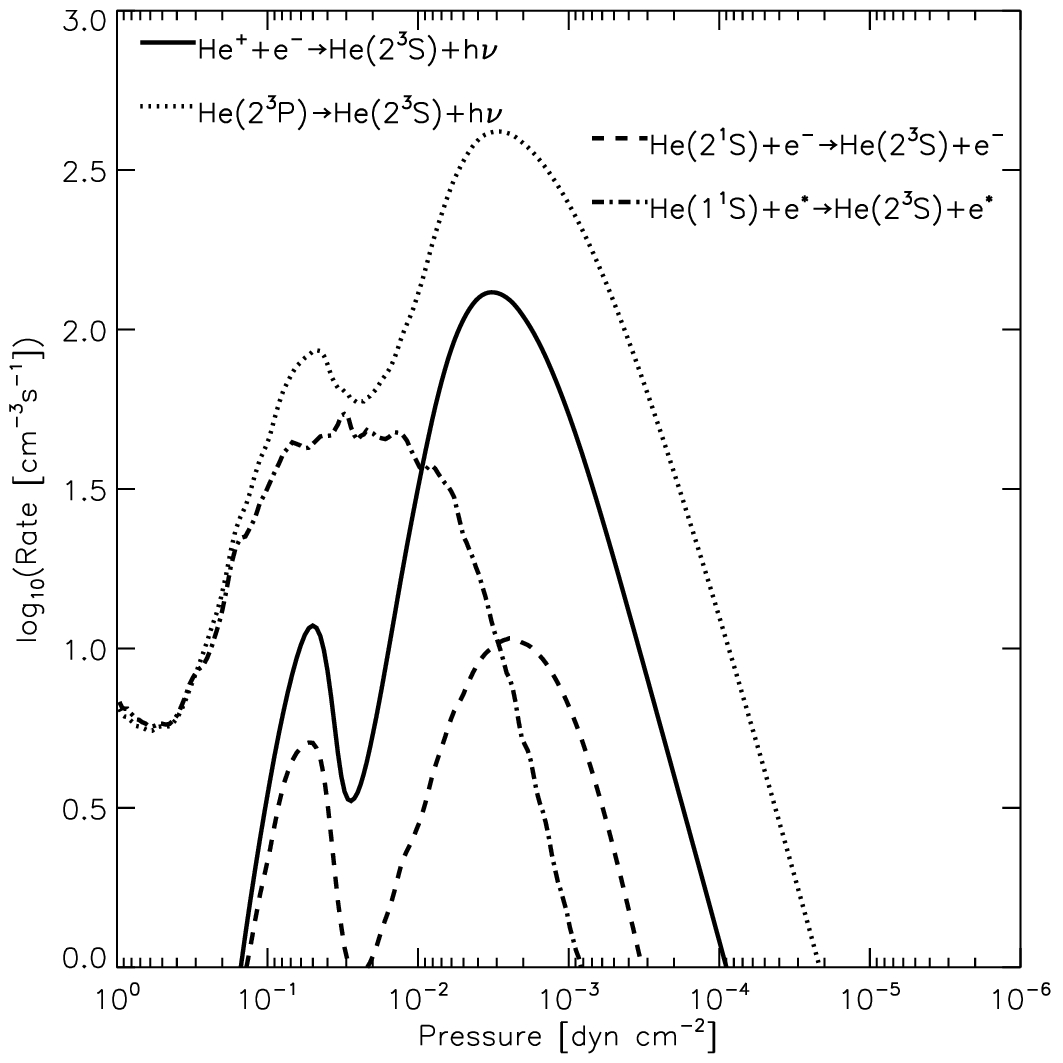}   
   \includegraphics[width=9.cm]{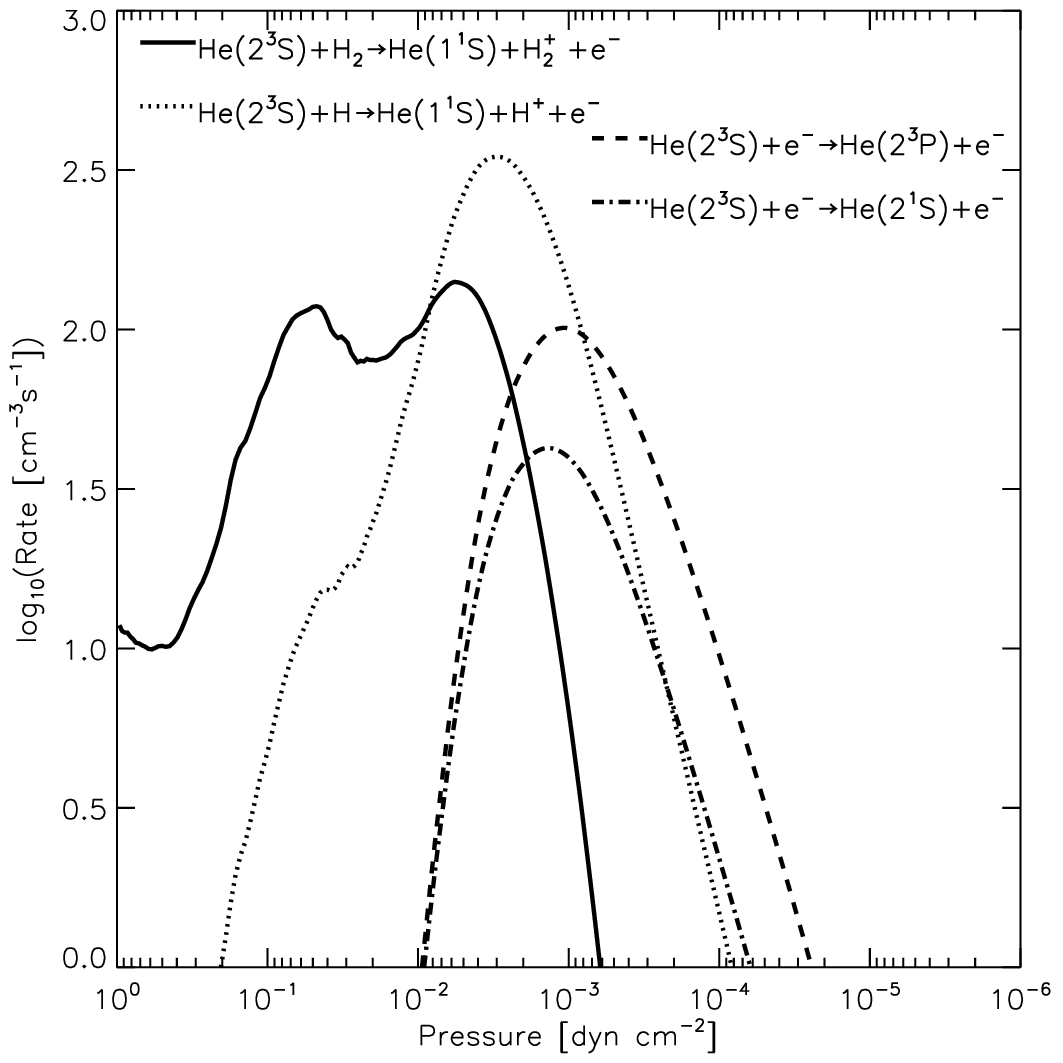}   
      \caption{\label{rates_h_2_3s_fig} From the MUS-b model.  
      Rates of the dominant He(2$^3S$) production (top) and destruction (bottom) processes; 
{\eminus} refers to thermal electrons, and $e^*$ to photoelectrons.      
}
\end{figure}

\section{Results}
 
To address Q1-Q3, we first ran 4 models of GJ 436 b. 
Two models use the MUSCLES SED, 
the other two use the X-Exoplanets SED. 
They are identified as MUS-a, MUS-b, X-Exo-a and X-Exo-b. 
Models with the label `-a' (`-b')  omit (include) the effect of photoelectrons. 
The MUSCLES SED describes a more strongly emitting star, thereby enhancing further the action of photoelectrons. 
That is the reason why we prioritize the MUS-a and -b models in the discussion below. 
\\

To address Q1, Fig. \ref{profiles_fig} (top) shows 
for the MUS-a and -b models 
the rates at which the primary and secondary electrons are released. The 
 dashed color lines refer to the 
 primaries, separated by process; the 
 dotted color lines refer to the secondaries, also separated by process; the  dashed black line refers to the primaries summed over all processes and, lastly, the
 solid black  line refers to the
primaries plus secondaries summed over all processes. The relative contribution of the secondaries becomes significant or dominant
at $p${$\gtrsim$}10$^{-2}$ dyn cm$^{-2}$ ($r$/$R_{\rm{p}}${$\lesssim$}1.07), where the main donors of primaries are {\htwo} and He and the mean energies of the primaries are 100-300 eV. \\

For comparison, Fig. \ref{profiles_fig} (bottom) shows the {\eminus}, {\htwo}, H, {\hp}, He, {\hep} and He($2^3S$) densities. 
Enhanced ionization by the secondaries leads to lower densities of neutrals and  higher densities of charged particles. 
The differences in the electron densities between the MUS-a and MUS-b models however never exceed 50{\%} and occur preferentially at $r$/$R_{\rm{p}}${$<$}1.1. 
At $r$/$R_{\rm{p}}${$\sim$}1.1, 
neutralization proceeds mainly 
through  (fast) dissociative recombination, {\hthreep}+{\eminus}{$\rightarrow$} 3H, {$\rightarrow$}{\htwo}+H, 
ensuring that the newly formed ions are rapidly replaced by neutrals and thus undoing the initial ionization step. 
It is likely that the mild atmospheric response to 
photoelectron-driven ionization 
seen in Fig. \ref{profiles_fig}
is specific to atmospheres in which molecules remain abundant. 
Foreseeably, if the atmospheric gas is in atomic form, neutralization will be less efficient because it must proceed through (slow) radiative recombination. 
\\

For reference, Fig. \ref{velotemp_fig} shows the temperature and velocity profiles for both the MUS-b and X-Exo-b models. 
They are generally lower for the latter, as expected from a weaker stellar irradiation. $T${$\sim$}5,000 K
at the peaks of the He(2$^3S$) density.
\\

To address Q2, Fig. \ref{rates_h_2_3s_fig} (top) shows for the MUS-b model the rates of the main He($2^3S$) production processes. 
Near the He($2^3S$) density peak at 
$p${$\sim$}1.5$\times$10$^{-3}$ dyn cm$^{-2}$ ($r$/$R_{\rm{p}}${$\sim$}1.21), 
radiative recombination dominates, either directly or indirectly (initially populating He($2^3P$), which decays promptly into He($2^3S$)).
Towards the model's lower boundary, photoelectron-driven excitation from He ground state dominates. The latter occurs either by direct excitation into He($2^3S$) or by excitation into and subsequent radiative decay from He($2^3P$). The borderline between photoelectron-driven excitation and radiative recombination as the dominant He(2$^3S$) production processes is at $p${$\sim$}10$^{-2}$ dyn cm$^{-2}$. The distinction is important because it defines the source of the He(2$^3S$) probed during transit. 
\\

Correspondingly, and to address Q3, Fig. \ref{rates_h_2_3s_fig} (bottom) shows the rates of the main He($2^3S$) loss processes. Penning ionization of {\htwo} towards the  model's lower boundary and of H near the He(2$^3S$) density peak set the He metastable's lifetime at  $p${$\gtrsim$}5$\times$10$^{-4}$ dyn cm$^{-2}$. At lower pressures, aided by temperatures {$\sim$}5,000 K, excitation into He(2$^1S$) and He(2$^3P$) in collisions with thermal electrons becomes competitive at removing He(2$^3S$). Most He(2$^3P$) formed this way will decay back into He(2$^3S$). Excitation into 
He(2$^1S$) ensures some singlet-triplet mixing, thus opening new possibilities for the ultimate removal of He(2$^3S$). 
The layers at $p${$\sim$}5$\times$10$^{-4}$ dyn cm$^{-2}$ (based on the MUS-a and MUS-b models, see below) are very important because they set the planet's size at the core of the {\hei} line. 
It follows that competition between He(2$^3S$)+H{$\rightarrow$}ionization and 
He(2$^3S$)+{\eminus}{$\rightarrow$}excitation
at removing He(2$^3S$) introduces a complex dependence on temperature and the H and {\eminus} densities in the formation of the line. 
We note that neither photoionization nor the photoelectron-driven processes (for excitation, deexcitation or ionization) seem to really affect the He(2$^3S$) loss at these low pressures. 
In the atmospheric layers that contribute to the in-transit spectrum, the lifetime of He(2$^3S$) remains $<$1 s, much shorter than its radiative lifetime. 
As a consequence, the He(2$^3S$) chemistry in the layers probed by in-transit spectroscopy is strongly coupled to the local gas conditions.
\\

The above discussion helps rationalize the differences in He(2$^3S$) densities between the -a and -b models.
The relative differences in the He(2$^3S$) population are large at $p${$\gtrsim$}10$^{-2}$ dyn cm$^{-2}$, where 
photoelectron-driven excitation
of He(1$^1S$) into He(2$^3S$) is strong. 
The differences in the He(2$^3S$) population
are barely {$\sim$}10$\%$ near the He(2$^3S$) density peak, where photoelectron effects are weak.
\\

\begin{figure}
   \centering
   \includegraphics[width=9.cm]{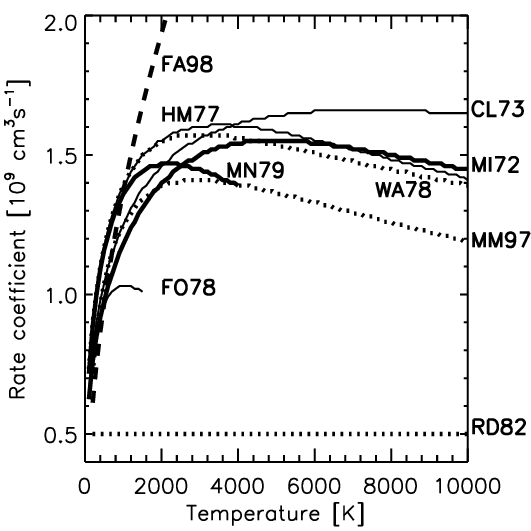}   
      \caption{\label{penning_kr_fig} Rate coefficients for total ionization through the processes He(2$^3S$)+H{$\rightarrow$}He(1$^1S$)+{\hp}+{\eminus}, {$\rightarrow$}{\hehp}+{\eminus}. 
      We calculated them  from the (scanned) cross sections reported in the quoted references over a range of temperatures commensurate with the range of energies at which they were reported. The calculations were performed by means of standard expressions, and assumed 
      a Maxwellian distribution for the relative velocities between the colliders. 
      Occasionally, we extrapolated the cross sections towards low energies with a $\propto${$E^{-0.2}$} law
      ($E$ is the center-of-mass kinetic energy)       
      and towards high energies with a $\propto${$E^{-1}$} law.
      References:  
      MI72, \citet{milleretal1972}; 
      CL73, \citet{cohenlane1973};  
      HM77, \citet{hickmanmorgner1977};      
      FO78, \citet{fortetal1978}; 
      MN79, \citet{morgnerniehaus1979}; 
      WA88, \citet{waibeletal1988}; 
      MM97, \citet{movremeyer1997}. 
      Also included is the value quoted by \citet{robergedalgarno1982} and the temperature-dependent expression utilized by \citet{fassiaetal1998}.
      The expression $k$ [cm$^3$s$^{-1}$]=10{$^{-9}$} $\exp(c/T + d_1 \ln{T} + d_2(\ln{T})^2 + d_3(\ln{T})^3)$ 
      ($c$=$-$8.64804E$+$01; 
       $d_1$=$-$2.86766E$-$01;
       $d_2$=$+$8.68445E$-$02; 
       $d_3$=$-$5.73001E$-$03) fits the rate coefficient based on the \citet{movremeyer1997} cross sections with errors {$<$}2{\%} from 200 to 10,000 K.
}
\end{figure}

\begin{figure}
   \centering
   \includegraphics[width=9.cm]{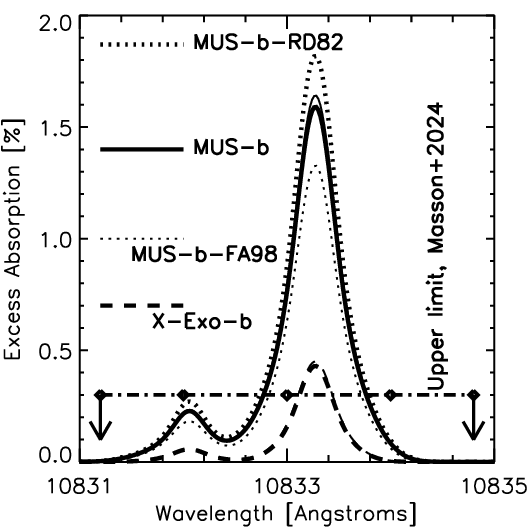}   
      \caption{\label{intransit_spectra_fig} Synthetic in-transit spectra based on the
      MUS-a, MUS-b, X-Exo-a, X-Exo-b, and MUS-b-RD82 and MUS-b-FA98 models (see text for details; thin and thick lines refer to -a and -b models).       
      The spectra are calculated following standard methods 
      \citep{garciamunozschneider2019}, with which we obtain the wavelength-dependent       
       effective radius $R_{\lambda}$. The excess absorption is defined as EA=($(R_{\lambda}/R_{\star})^2${$-$}$(R_{\rm{p}}/R_{\star})^2$){$\times$}100, with $R_{\rm{p}}$ being the planet's optical radius and $R_{\star}$ the stellar radius. 
       Also shown is the non-detection upper limit reported by \citet{massonetal2024}. \citet{nortmannetal2018} and \citet{guilluyetal2024} report upper limits of 0.41 and 0.42{\%}, respectively.
}
\end{figure}

\section{Sensitivity analysis. In-transit spectrum}

The preceding analysis confirms the importance of Penning ionization 
for the loss of the He metastable
in He(2$^3S$)+H collisions, possibly together with excitation into higher-energy states through He(2$^3S$)+{\eminus} collisions. 
It is instructive to compare the simulations based on the rate coefficient for Penning ionization calculated by us  (Table \ref{penning_table}), 
which complies with the available experimental and numerical evidence, 
with simulations based on other prescriptions of the rate coefficient.\\

The MUS-b-RD82 model is identical to the MUS-b model except that it uses a $T$-independent
rate coefficient of 5$\times$10$^{-10}$ cm$^3$s$^{-1}$ for the total of Penning and associative ionization
He(2$^3S$)+{H}{$\rightarrow$}{He(1$^1S$)}+H$^+$+{\eminus}, {$\rightarrow$}{HeH}$^+$+{\eminus}. This is the value quoted by \citet{robergedalgarno1982}, who in turn quote \citet{fortetal1978}, \citet{neynabertang1978} and \citet{morgnerniehaus1979}. 
We inspected the original references to track the discrepancy by more than a factor of 2 between 
the \citet{robergedalgarno1982} rate coefficient 
and ours.
It appears that the cross sections reported in the aforementioned references are generally consistent with the \citet{movremeyer1997} cross sections that we used, and with 
later cross section determinations such as those by \citet{waibeletal1988}.
Figure \ref{penning_kr_fig} elaborates further on the comparison with these and other sources.
As a consequence, the moderate differences between cross sections cannot explain the large discrepancy in the rate coefficients. 
We speculate that the
\citet{robergedalgarno1982} value was determined at very low temperature (our calculations yield 
{$\sim$}7$\times$10$^{-10}$ cm$^3$s$^{-1}$ at 100 K) or that the authors mistakenly adopted the cross sections for He(2$^3S$)+{D}{$\rightarrow$}ionization from \citet{neynabertang1978}. 
 The MUS-b-FA98 model adopts for Penning ionization of H atoms the {$\propto$}$T^{1/2}$ law noted earlier. No details are given in 
\citet{fassiaetal1998} to motivate this extrapolation law.\\

Figure \ref{intransit_spectra_fig} presents synthetic
in-transit spectra of the {\hei} line at 1.08 $\mu$m for the MUS-a, MUS-b, X-Exo-a, X-Exo-b, MUS-b-RD82 
and MUS-b-FA98 models of GJ 436 b. They were prepared with routines developed in  \citet{garciamunozschneider2019}, and expressed in the form of excess absorptions. 
The line optical properties are from the NIST  database 
\citep{kramida2010,kramidaetal2018}.
The figure includes the upper limit 
on the excess absorption
inferred by \citet{massonetal2024} from their non-detection of the {\hei} line at GJ 436 b. The strength of the in-transit signal is connected to the capacity of the stellar irradiation to drive a vigorous outflow and, therefore, to the planet's mass loss rate. We obtain (integrated over a solid angle of 4$\pi$) mass loss rates $\sim$2.3$\times$10$^{10}$ g/s for the simulations based on the MUSCLES SED and 
$\sim$7.4$\times$10$^{9}$ g/s for the simulations based on the X-Exoplanets SED. Taking into account the effect of photoelectrons reduces the mass loss rates by $\sim$5-10{\%}. 
\\

The MUS-a and -b spectra are barely distinguishable. 
Similarly, the X-Exo-a and -b spectra are nearly identical. 
The reason is that the effects of photoelectrons on He(2$^3S$) are confined to atmospheric layers that are too deep and contribute little to the in-transit spectra. In contrast, the He(2$^3S$) produced by radiative recombination forms at higher altitudes and effectively sets the stellar area masked by the {\hei} line during transit. 
It is difficult to anticipate how general this finding is, and whether the He(2$^3S$) formed from photoelectron-driven excitation makes a bigger difference at other planets. We will test this idea in future work by simulating additional planet-star systems.\\

The
comparison of the MUS-b and X-Exo-b spectra demonstrates the importance of the adopted SED. 
Whereas the MUS-a and -b spectra are clearly inconsistent with the published non-detections, the X-Exo-a and -b spectra are notably closer to the reported upper limits \citep{nortmannetal2018, guilluyetal2024,massonetal2024}. 
The choice of SED may partly explain the problems encountered by past models to explain the non-detections for GJ 436 b \citep{oklopcichirata2018,
dossantosetal2022,rumenskikhetal2023}. Taken at face value, the better match obtained with the X-Exoplanets SED suggests that this may be a closer representation of the true stellar SED. This possibility could in principle be tested only with an empirical representation of the SED.
\\

Lastly, we describe the comparison between the 
MUS-b, MUS-b-RD82 and MUS-b-FA98 models. 
The He(2$^3S$) peak densities (not shown)  differ by a factor of {$\sim$}3 between the MUS-b-RD82 and MUS-b-FA98 models, somewhat less than the factor of 6 in the adopted rate coefficients at $T$=5,000 K. The corresponding difference in the excess absorption of the in-transit spectra (Fig. \ref{intransit_spectra_fig}) is smaller, but  larger than the errors often quoted in this type of measurements. As expected, the
predicted in-transit spectrum based on our recommended rate coefficient for Penning ionization of H atoms sits somewhere in between. The comparison clearly conveys that significant biases can be introduced by describing some of the key Ch-C-R processes in ways that are not physically motivated.
\\

\section{Summary}

We propose a revised network of chemical-collisional-radiative processes for modelling 
He(2$^3S$) in exoplanet atmospheres. 
It generalizes past approaches in that it enables the simulation of outflows in which {\htwo} remains undissociated to high altitudes, a valuable addition towards the study of Neptune- and sub-Neptune-sized planets. 
In our simulations of GJ 436 b, 
we identify Penning ionization of H atoms as a key process at removing the high-altitude He(2$^3S$) that sets the planet's size at the core of the {\hei} line at 1.08 $\mu$m. We calculated the rate coefficient for this process from published cross sections. It differs by factors of a few with respect to other rate coefficient prescriptions found in the literature. A sensitivity analysis shows that the misrepresentation of Penning ionization significantly biases the strength of the {\hei} line at its core. More generally, the misrepresentation of this process may have implications on the He/{\htwo} abundance ratio constrained from observation-model comparisons of other exoplanets. 
\\

Large changes in ionization rates driven by photoelectrons do not lead to similarly large changes in the densities of the charged particles. This may be a property of atmospheres in which the background gas remains in molecular form where most of the secondary electrons are released. The idea merits further attention by testing it at other planet-star systems.  
At GJ 436 b, 
photoelectron-driven excitation from He ground state into He(2$^3S$) dominates the 
production of He metastable
at $p${$\gtrsim$}10$^{-2}$ dyn cm$^{-2}$. 
These layers are however too deep to make a difference in the in-transit spectrum. 
Our finding does not rule out however that photoelectron-driven processes may have a stronger effect in other atmospheres, especially those in which the transition from molecules to atoms occurs in deeper layers.
\\

Lastly, our investigation reveals that the He(2$^3S$) population at GJ 436 b is very sensitive to the adopted stellar SED. We tested two SED and found that the simulated outflows result in very different in-transit signals. The simulation based on the weakest of the two SED is nearly consistent with the non-detections of the {\hei} line at 1.08 $\mu$m from three independent groups. 
\\

\section*{Data Availability}

Fits to the rate coefficients can be found through the 
\href{https://doi.org/10.5281/zenodo.15453858}{link}.
\\

\begin{acknowledgements}
Thanks are due to I. Bray, 
D. De Fazio, 
M. G\'en\'evriez, 
J. Loreau, 
S. Nahar and 
X. Urbain for feedback on different aspects of the collisional-radiative processes in helium.
\end{acknowledgements}

   \bibliographystyle{aa} 
   \bibliography{mybiblio}

\clearpage

\begin{appendix}

\section{Helium network}

For reference, Tables \ref{states_table}--\ref{nonthermalchannels_table} list information that describes the species and
Ch-C-R processes specific to helium
in our model.
In these tables and throughout the paper, we follow the notation $X$E$+Y$=$X${$\times$}10$^{+Y}$.
When appropriate, the rate coefficients have been fitted to expressions of the form $a T^b \exp{(c/T)} \Upsilon (T)$, with $\ln{\Upsilon}$=$\sum_{t=0}^3 d_t (\ln{T})^t$ and where $a$, $b$, $c$ and $d_t$ are fit coefficients.
Only 5 of the 7 fit coefficients are independent. 
A SI file contains the details of the expressions.
\\

\begin{table}[H]
\caption{\label{states_table} Atom model for helium.}
\centering                          
\begin{tabular}{c c c c}        
\hline                 
Index, $i$ & {\hei} & {\heii} & {\hehp}  \\    
\hline
  & He/He(1$^1S$) & {\hep}/{\hep}(1$^2S$) & {\hehp} \\
1 & 1    & 2  &  \\
  & +00.0000 & +24.5874 & +14.0269\\
\hline
  & He(2$^3S$) &    \\
2 & 3    &   \\
  & +19.8195   &  \\
\hline

  & He(2$^1S$) &    \\
3 & 1    &   \\
  & +20.6157   &  \\
\hline

  & He(2$^3P$) &    \\
4 & 9    &   \\
  & +20.9641   &  \\
\hline

  & He(2$^1P$) &    \\
5 & 3    &   \\
  & +21.2181   &  \\
\hline

\end{tabular}
\tablefoot{For {\hei} and {\heii}:  
denominations, statistical weights and energies [eV] of the states adopted in the network;  
borrowed from the NIST database 
\citep{kramida2010,kramidaetal2018}.
For {\hehp}: formation energy; borrowed from the KIDA database \citep{wakelametal2012}.}             
\end{table}

\begin{table}[H]
\caption{\label{radrecom_table} Effective rate coefficients $\alpha_{i,\rm{eff}}$ [cm$^3$s$^{-1}$] at selected temperatures for radiative recombination, 
{\hep}+{\eminus}{$\rightarrow$}{He}($i$)+$h\nu$.}
\centering                        
\begin{tabular}{l r c c c c}        
\hline                 
   &  &    \multicolumn{4}{c}{$T$ [K]}    \\
   \cline{3-6}
$i$    & \multicolumn{1}{c}{[{\AA}]}    & 500 &  2,000     &    5,000     &    10,000    \\
\hline
1      &     504.3     & 6.76E$-$13 & 3.38E$-$13 & 2.15E$-$13 & 1.53E$-$13    \\
2      &    2600.4     & 3.87E$-$13 & 1.71E$-$13 & 9.85E$-$14 & 6.40E$-$14    \\
3      &    3121.7     & 1.35E$-$13 & 5.87E$-$14 & 3.28E$-$14 & 2.06E$-$14    \\
4      &    3421.9     & 1.46E$-$12 & 5.33E$-$13 & 2.61E$-$13 & 1.47E$-$13    \\
5      &    3679.8     & 4.49E$-$13 & 1.60E$-$13 & 7.64E$-$14 & 4.19E$-$14    \\

\hline
\end{tabular}
\tablefoot{Also indicated, the wavelength thresholds for continuum emission. 
Index $i$ refers to the states listed in Table \ref{states_table}.
}         

\end{table}

\begin{table}[H]
\caption{\label{Aji_table} Adopted transition probabilities $A_{ji}$ [s$^{-1}$] between He states.}
\centering                          
\begin{tabular}{l l l l l}        
\hline                 
\multicolumn{5}{c}{He($j$){$\rightarrow$}He($i$)+{$h\nu$}} \\
\hline
$j${$\rightarrow$}{$i$} & 1 & 2  & 3 & 4      \\    
\hline
2 & 1.27E$-$04 &            &              &  \\
3 & 5.09E+01   & 0.00E+00 &              &  \\
4 & 1.76E+02   & 1.02E+07 & 2.96E$-$02 &  \\
5 & 1.79E+09   & 1.44E+00 & 1.97E+06   & 0.00E+00 \\

\hline

\end{tabular}
\tablefoot{Indices $i$ and $j$ refer to the states listed in Table \ref{states_table}.}
\end{table}

\begin{figure}[H]
   \centering
   \includegraphics[width=10.cm]{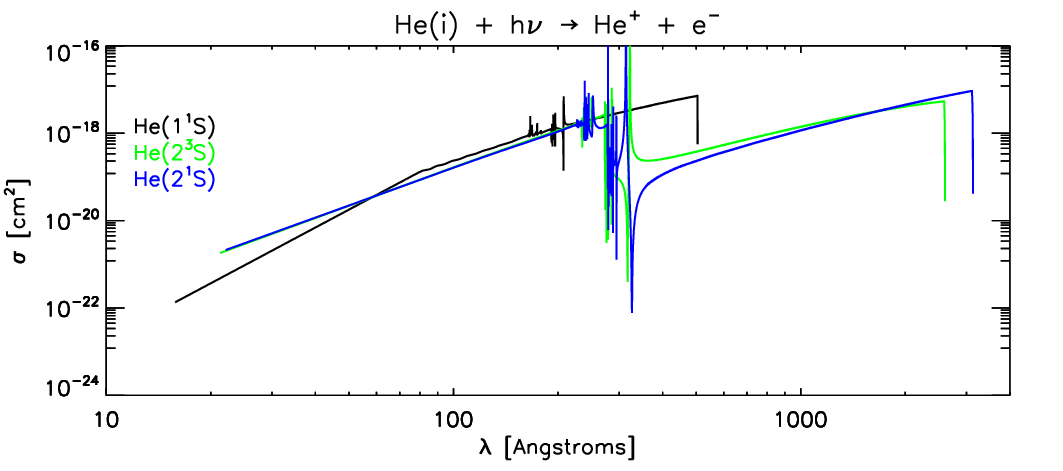}
      \caption{\label{xs_he_fig} He photoionization cross sections adopted in the network. From the NORAD database \citep{nahar2010,nahar2020}. }
\end{figure}

\begin{figure}[H]
\centering
\includegraphics[width=10cm]{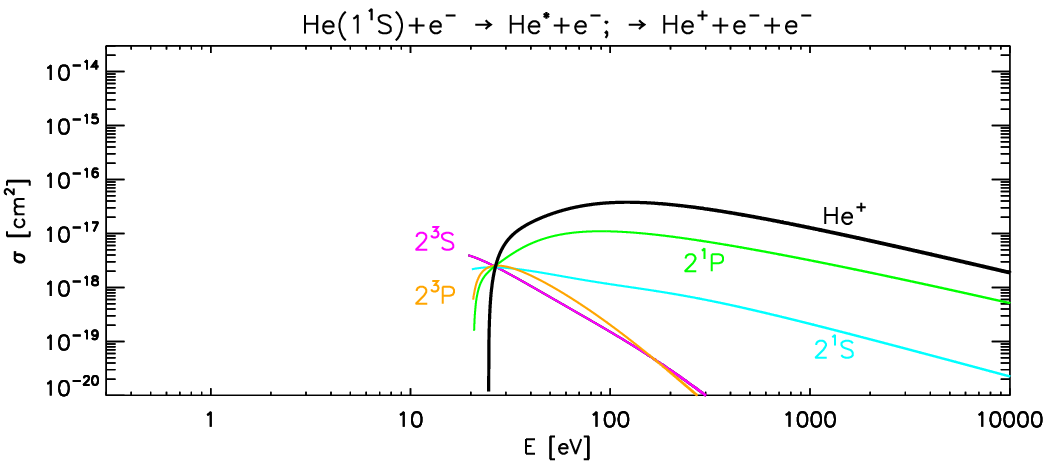} \\ 
\vspace{-0.2cm}
\includegraphics[width=10cm]{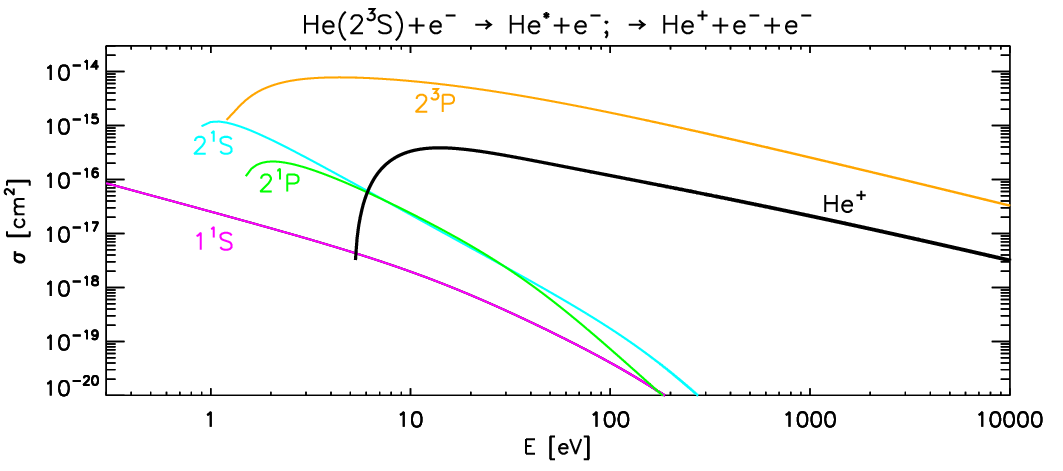} \\ 
\vspace{-0.2cm}
\includegraphics[width=10cm]{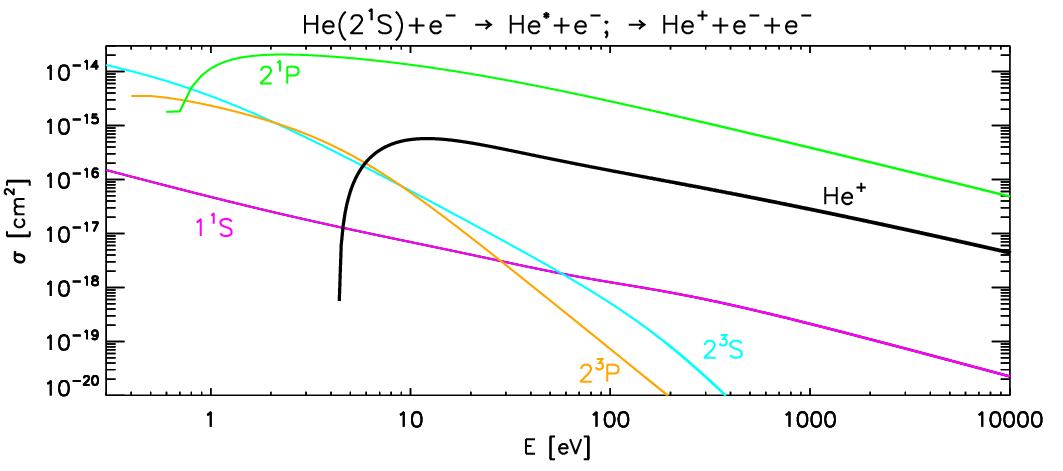} \\ 
\caption{Excitation, deexcitation and ionization 
cross sections for collisions of electrons of the specified energies with He atoms. Based on \citet{ralchenkoetal2008}. 
    } 
    \label{electronimpactxs_fig}
\end{figure}

\begin{table*}
\caption{\label{em-collexc_table} 
Rate coefficients at selected temperatures for (de)excitation in collisions with thermal 
electrons, He($i$)+{\eminus}{$\rightarrow$}He($j$)+{\eminus}.}
\centering                          
\begin{tabular}{l l l l l l}        
\hline                 
\multicolumn{6}{c}{$T$= 500 K} \\
\hline
$i${$\rightarrow$}{$j$} & 1 & 2  & 3 & 4  & 5    \\    
\hline
1 &            & 1.22E$-$208 & 6.79E$-$217 & 7.47E$-$221 & 1.03E$-$223 \\
2 & 2.42E$-$09 &             & 6.36E$-$16  & 6.66E$-$19  & 1.88E$-$22  \\
3 & 4.25E$-$09 & 2.02E$-$07  &             & 8.69E$-$11  & 1.11E$-$13  \\
4 & 1.69E$-$10 & 7.65E$-$08  & 3.13E$-$08  &             & 1.53E$-$10  \\
5 & 2.56E$-$10 & 2.35E$-$08  & 4.38E$-$08  & 1.67E$-$07  &             \\
\hline
\multicolumn{6}{c}{$T$= 2,000 K} \\
\hline
$i${$\rightarrow$}{$j$} & 1 & 2  & 3 & 4  & 5    \\    
\hline
1 &            & 1.02E$-$58  & 4.88E$-$61 & 2.85E$-$62  & 3.21E$-$63 \\
2 & 3.00E$-$09 &             & 1.02E$-$09 & 4.93E$-$10  & 9.40E$-$12  \\
3 & 4.34E$-$09 & 3.11E$-$07  &            & 3.18E$-$08  & 1.77E$-$08  \\
4 & 2.13E$-$10 & 1.26E$-$07  & 2.67E$-$08 &             & 1.28E$-$08  \\
5 & 3.14E$-$10 & 3.14E$-$08  & 1.95E$-$07 & 1.68E$-$07  &             \\
\hline

\multicolumn{6}{c}{$T$= 5,000 K} \\
\hline
$i${$\rightarrow$}{$j$} & 1 & 2  & 3 & 4  & 5    \\    
\hline
1 &            & 8.04E$-$29  & 6.20E$-$30  & 1.47E$-$30  & 4.34E$-$31 \\
2 & 2.54E$-$09 &             & 1.48E$-$08  & 3.79E$-$08  & 1.20E$-$09  \\
3 & 3.73E$-$09 & 2.81E$-$07  &             & 8.42E$-$08  & 2.86E$-$07  \\
4 & 2.21E$-$10 & 1.80E$-$07  & 2.10E$-$08  &             & 2.69E$-$08  \\
5 & 3.52E$-$10 & 3.08E$-$08  & 3.87E$-$07  & 1.45E$-$07  &             \\
\hline

\multicolumn{6}{c}{$T$= 10,000 K} \\
\hline
$i${$\rightarrow$}{$j$} & 1 & 2  & 3 & 4  & 5    \\    
\hline
1 &            & 5.86E$-$19  & 1.24E$-$19  & 5.27E$-$20  & 2.31$-$20 \\
2 & 1.90E$-$09 &             & 2.81E$-$08  & 1.90E$-$07  & 5.35$-$09  \\
3 & 3.05E$-$09 & 2.12E$-$07  &             & 9.79E$-$08  & 8.25$-$07  \\
4 & 2.15E$-$10 & 2.40E$-$07  & 1.63E$-$08  &             & 2.99$-$08  \\
5 & 3.80E$-$10 & 2.71E$-$08  & 5.53E$-$07  & 1.20E$-$07  &             \\
\hline

\end{tabular}
\tablefoot{Indices $i$ and $j$ refer to the states listed in Table \ref{states_table}.}            
\end{table*}

\begin{table*}
\caption{\label{penning_table} 
Rate coefficients [cm$^3$s$^{-1}$] at selected temperatures for 
loss processes of metastable He in collisions with neutrals.
}
\centering                        
\begin{tabular}{c c c c c c}        
\hline                 

   &    \multicolumn{4}{c}{$T$ [K]}   &   \\
   \cline{2-5}
Channel    & 500 &  2,000     &    5,000     &    10,000  &  Ref. \\
\hline

He(2$^3S$)+{H}{$\rightarrow$}{He(1$^1S$)}+H$^+$+{\eminus}, {\hehp}+{\eminus}  & 1.02E$-$09 &  1.32E$-$09 &  1.35E$-$09  & 1.27E$-$09 & \citet{movremeyer1997} \\
He(2$^1S$)+{H}{$\rightarrow$}{He(1$^1S$)}+H$^+$+{\eminus}, {\hehp}+{\eminus}  & 2.54E$-$09 &  2.89E$-$09 &  2.95E$-$09  & 2.96E$-$09 & \citet{movreetal1994} \\

\hline

He(2$^3S$)+{\htwo}{$\rightarrow$}{He(1$^1S$)}+{\htwop}+{\eminus}, H+{\hehp}+{\eminus} & 8.94E$-$11 & 6.48E$-$10  & 1.48E$-$09 &  2.54E$-$09 & \citet{cohenlane1977} \\
He(2$^1S$)+{\htwo}{$\rightarrow$}{He(1$^1S$)}+{\htwop}+{\eminus}, H+{\hehp}+{\eminus} & 1.50E$-$10 & 9.44E$-$10  & 1.85E$-$09 &  2.79E$-$09 & \citet{cohenlane1977} \\

\hline

He(2$^1S$)+{He}{$\rightarrow$}{He}+{He}+$h \nu$  & 6.52E$-$15 & 1.56E$-$14 & 1.95E$-$14 & 2.16E$-$14 & \citet{zygelman1991}\\

\hline

\end{tabular}
\end{table*}

\begin{table*}
\caption{\label{hehp_table} Rate coefficients [cm$^3$s$^{-1}$] 
at selected temperatures
for some processes involving {\hehp}.
}         
\centering                        
\begin{tabular}{c c c c c c}        
\hline                 

   &    \multicolumn{4}{c}{$T$ [K]}   &   \\
   \cline{2-5}
Channel    & 500 &  2,000     &    5,000     &    10,000  &  Ref. \\
\hline
{\hehp}+{\eminus}{$\rightarrow$}{He}+H &  2.57E$-$08 & 1.70E$-$08 &  1.29E$-$08 &  1.05E$-$08 & \citet{florescu-mitchellmitchell2006} \\
{\hehp}+{\htwo}{$\rightarrow$}{\hthreep}+He & 1.26E$-$09 & 1.26E$-$09 & 1.26E$-$09 & 1.26E$-$09 & \citet{orient1977} \\
{\hehp}+{H}{$\rightarrow$}{\htwop}+He & 1.27E$-$09 & 1.73E$-$09 & 2.05E$-$09 & 2.33E$-$09 & \citet{defazio2014} \\

\hline

{\htwop}+{He}{$\rightarrow$}{\hehp}+H & 4.39E$-$16 & 1.04E$-$11 & 7.83E$-$11 & 1.53E$-$10 & \citet{black1978} \\

\hline

\end{tabular}
\end{table*}

\begin{table*}
\caption{\label{chargeexchange_table} Rate coefficients
[cm$^3$s$^{-1}$] at selected temperatures for charge exchange.
}         
\centering                        
\begin{tabular}{c c c c c c}        
\hline                 

   &    \multicolumn{4}{c}{$T$ [K]}   &   \\
   \cline{2-5}
Channel    & 500 &  2,000     &    5,000     &    10,000  &  Ref. \\
\hline
{\hep}+{\htwo}{$\rightarrow$}{He}+{\hp}+H, $\dagger$ & 3.00E$-$14 &  3.00E$-$14 &  3.00E$-$14  & 3.00E$-$14 & \citet{schaueretal1989} \\
{\hep}+{H}{$\rightarrow$}{\hp}+He, $\dagger$ & 2.00E$-$15 &  2.00E$-$15 &  2.00E$-$15  & 2.00E$-$15 & \citet{courtneyetal2021} \\

\hline

{\hp}+He(2$^3S$){$\rightarrow$}{\hep}+H & 1.37E$-$20 & 3.25E$-$16 & 3.42E$-$15  & 1.89E$-$13 & \citet{loreauetal2018} \\
{\hp}+He(2$^1S$){$\rightarrow$}{\hep}+H & 7.83E$-$16 & 4.28E$-$15 & 2.38E$-$13  & 1.05E$-$11 & \citet{loreauetal2018} \\
\hline

\end{tabular}
\tablefoot{$\dagger$: Assumed $T$-independent.}
\end{table*}

\begin{table*}
\caption{\label{nonthermalchannels_table}Photoelectron-driven processes in the network.  }         
\centering                        
\begin{tabular}{l c c l c}        
\hline                 

Indices & \multicolumn{3}{c}{Channel} & Threshold [eV] \\

\hline
(1,2) & He(1$^1S$)+$e'$    & {$\rightarrow$} &  He(2$^3S$)+$e''$ & 19.8195    \\
(1,3) &                    & {$\rightarrow$} &  He(2$^1S$)+$e''$ & 20.6157   \\
(1,4) &                    & {$\rightarrow$} &  He(2$^3P$)+$e''$ & 20.9641   \\
(1,5) &                    & {$\rightarrow$} &  He(2$^1P$)+$e''$ & 21.2181   \\
(1,$\infty$) &             & {$\rightarrow$} &  {\hep}(1$^2S$)+$e''$+$e'''$ &  24.5874  \\
\hline

(2,1) & He(2$^3S$)+$e'$    & {$\rightarrow$} &  He(1$^1S$)+$e''$ &  0  \\
(2,3) &                    & {$\rightarrow$} &  He(2$^1S$)+$e''$ &  0.7962  \\
(2,4) &                    & {$\rightarrow$} &  He(2$^3P$)+$e''$ &  1.1446  \\
(2,5) &                    & {$\rightarrow$} &  He(2$^1P$)+$e''$ &  1.3986  \\
(2,$\infty$) &             & {$\rightarrow$} &  {\hep}(1$^2S$)+$e''$+$e'''$ & 4.7679   \\
\hline

(3,1) & He(2$^1S$)+$e'$    & {$\rightarrow$} &  He(1$^1S$)+$e''$ &  0  \\
(3,2) &                    & {$\rightarrow$} &  He(2$^3S$)+$e''$ &  0  \\
(3,4) &                    & {$\rightarrow$} &  He(2$^3P$)+$e''$ &  0.3484 \\
(3,5) &                    & {$\rightarrow$} &  He(2$^1P$)+$e''$ &  0.6024  \\
(3,$\infty$) &             & {$\rightarrow$} &  {\hep}(1$^2S$)+$e''$+$e'''$ & 3.9717   \\
\hline

\end{tabular}
\tablefoot{Indices ($i,j$) refer to the initial ($i$) and final ($j$) states in {\hei} (see Table \ref{states_table}), and $\infty$ refers to {\heii}.}
\end{table*}

\end{appendix}

\end{document}